\documentclass[pre,preprintnumbers,amsmath,amssymb]{revtex4}
\usepackage{graphicx}
\usepackage{latexsym}
\usepackage{color}

\begin{document}

\title{Traversable asymptotically flat wormholes in Rastall gravity}
\author{H. Moradpour$^1$\footnote{hn.moradpour@gmail.com}, N. Sadeghnezhad$^1$\footnote{nsadegh@riaam.ac.ir}, S. H. Hendi$^2$\footnote{hendi@shirazu.ac.ir}}
\address{$^1$ Research Institute for Astronomy and Astrophysics of Maragha (RIAAM), P.O. Box 55134-441, Maragha, Iran\\
$^2$ Physics Department and Biruni Observatory, Shiraz University,
Shiraz 71454, Iran}

\begin{abstract}
\noindent There are some gravitational theories in
which the ordinary energy-momentum conservation law is not valid in
the curved spacetime. Rastall gravity is one of the known theories
in this regard which includes a non-minimal coupling between
geometry and matter fields. Equipped with the basis of such theory,
we study the properties of traversable wormholes with flat
asymptotes. We investigate the possibility of exact solutions by a
source with the baryonic matter state parameter.
Our survey indicates that Rastall theory has considerable effects on
the wormhole characteristics. In addition, we study
various case studies and show that the weak energy condition may be
met for some solutions. We also give a discussion regarding to
traversability of such wormhole geometry with phantom sources.

\end{abstract}

\maketitle

\section{Introduction}
$\mathcal{W}$ormholes as the backbone of interstellar travels
\cite{Wheeler1,Wheeler2}, should be traversable
\cite{Thorne1,Thorne2}. Some primary solutions for traversable
wormholes have also been derived by M. Visser \cite{Visser1}. It is
also argued that a phantom energy may support the traversable
wormholes \cite{Sushkov,Lobo2005,Lobo20051}. Moreover, it is shown
that wormholes and black holes are convertible structures and in
fact, their physics are so close to each other
\cite{Hayward,Kardashev,Kuhfittig,Sushkov-Zaslavskii,Cai1,Cai2,ref1}.
These structures are also studied in modified theories of gravity
\cite{prds,prds1} such as the braneworld scenario
\cite{braneb,Lobo2}, conformal Weyl gravity \cite{Lobo1}, the $f(R)$
gravity \cite{Oliveira,Sajadi}, and the curvature-matter coupling
framework \cite{Garcia2,Garcia3} (for a detailed review see
\cite{loboaip}). In addition, one may use cut-and-paste method to
construct a thin-shell wormhole
which needs an exotic matter that violates the null
energy condition. Following this surgical
technique, various thin-shell wormhole solutions
have been addressed in the context of different gauge-gravity
theories
\cite{cut-paste1,cut-paste2,cut-paste3,cut-paste4,cut-paste44,cut-paste5,cut-paste6,cut-paste7,cut-paste8,cut-paste9,cut-paste10,cut-paste11,cut-paste12}.

In the curvature-matter theory of gravity \cite{cmc,cmc1,cmc2},
while the divergence of energy-momentum tensor is not always zero,
geometry and matter fields are coupled to each other in a
non-minimal way. Indeed, quantum effects in curved spacetimes such
as the particle production process
\cite{motiv2,motiv1,motiv4,motiv3,rw1} may motivate us to
consider non-divergenceless energy-momentum sources and thus modify
the general relativity theory. On the other hand, there is another
modification to the Einstein's theory proposed by P. Rastall
\cite{rastall}, which also couples the geometry to
the matter fields in a non-minimal way \cite{prd,smal,cosmos3}. As
it has been argued by Rastall \cite{rastall}, we only test the
energy-momentum conservation law in our laboratory and thus the flat
space. Indeed, to generalize the condition of null covariant
derivative of energy-momentum tensor from flat spacetime to the
curved spacetime is the simplest assumption to get general
relativity. Therefore, it is not forbidden to relax this condition
which leads to a modified general relativity theory \cite{rastall}.

Although the Rastall field equations are more simple than those of
the curvature-matter theory, it is in agreement with some
observational data on the Hubble parameter and the age of the
universe \cite{al1} meaning that it could be free of the entropy
and age problems of the standard cosmology \cite{al3}. In
addition, the Rastall theory leads to better consistency with the
observational data of the matter dominated era against the
Einstein field equations \cite{al2}. Observational data on the
helium nucleosynthesis also supports this theory \cite{al6}. More
studies on the cosmological features of the Rastall theory
including its consistency with various cosmic eras can be found in
\cite{prd,al5,al55}. Finally, it is worthwhile mentioning that
this theory provides an appropriate platform to investigate the
gravitational lensing \cite{al4,al7}. In addition, abelian-Higgs
strings in a phenomenological Rastall model have been analyzed in
Ref. \cite{arXiv:1407.3849}. Moreover, Rastall gravity has been
investigated in the context of the G\"{o}del-type universe with a
perfect fluid matter \cite{Godel} and it was shown that the
geodesics of particles does not altered.

Moreover, similar to the curvature-matter theory of gravity, in
the Rastall theory the divergence of energy-momentum tensor does
not always vanish in the curved spacetime \cite{rastall}, and
therefore, the energy-momentum conservation law is not always
valid. In fact, the curvature-matter theory is a kind of $f(R)$
gravity, in which matter and geometry are coupled to each other in
a non-minimal way \cite{cmc,cmc1,cmc2}, and its lagrangian differs
from that of the Rastall theory \cite{smal,cosmos3}. As we have
mentioned, the wormholes structures are addressed in the
curvature-matter coupling framework
\cite{Garcia2,Garcia3,loboaip}. Therefore, since the Rastall
theory differs from the curvature-matter coupling framework
\cite{cmc,cmc1,cmc2,smal,cosmos3}, it is useful to investigate the
structure of wormholes in this framework in order to get new
aspects of wormholes, Rastall theory and in fact, the effects of
considering a source (an energy-momentum tensor) with non-zero
divergence on the wormholes and spacetime structures.

Here, we will introduce some traversable asymptotically flat
wormholes in the Rastall framework and study their physical
properties. Moreover, we are interested in studying the effects of
considering an energy-momentum tensor, supporting the background,
with non-zero divergence on the wormhole structures and their
properties. In order to achieve such goal, we first review some
properties of the Rastall theory, and then try to get the energy
conditions in the Rastall theory. Besides, taking the Newtonian
limit, we derive a dimensionless parameter for the Rastall theory
which helps us in simplifying and classifying calculations in this
theory, meanwhile, some mathematical properties of traversable
asymptotically flat wormholes are also addressed. In addition, we
study some physical properties of the energy-momentum tensor
supporting the mentioned geometry in the Rastall theory. Our study
shows that the wormhole parameters are affected by the parameters of
Rastall theory.

The paper is organized as follows. In the next section, we review
the Rastall theory and point out some of its mathematical and
physical properties. In section ($\textmd{III}$), we address some
mathematical properties of asymptotically traversable wormholes.
Sections ($\textmd{IV}$) and ($\textmd{V}$) include some examples
for the traversable asymptotically flat wormholes in the Rastall
theory. We also study the properties of energy-momentum tensor,
supporting the geometry in this theory, as well as the relation
between the Rastall's and wormhole's parameters throughout the forth
and fifth sections. The last section is devoted to the summary.
Units of $c=\hbar=1$ are considered in this paper.
\section{A brief review on the Rastall theory}
Rastall questioned the validity of the energy-momentum
conservation law in the four dimensional spacetime \cite{rastall}.
His hypothesis ($T^{\mu \nu}_{\ \ ;\mu}\neq0$) leads to a
modification to the Einstein field equations in agreement with
various observational data
\cite{al1,al3,al2,al6,al5,al55,al4,al7}. Based on the Rastall's
theory \cite{rastall}, if the spacetime is filled by a source with
$T^{\mu}_{\nu}$, then
\begin{eqnarray}\label{rastal}
T^{\mu \nu}_{\ \ ;\mu}=\lambda R^{,\nu},
\end{eqnarray}
where $R$ and $\lambda$ are the Ricciscalar of the spacetime and
Rastall parameter, respectively. This equation leads to
\cite{rastall,cosmos3,smal,prd,al1,al3,al2,al6}
\begin{eqnarray}\label{r1}
G_{\mu \nu}+\kappa\lambda g_{\mu \nu}R=\kappa T_{\mu \nu},
\end{eqnarray}
which can finally be written as
\begin{equation}\label{ein}
G_{\mu \nu}=\kappa S_{\mu\nu},
\end{equation}
where $\kappa$ is the Rastall gravitational coupling constant, and
$S_{\mu\nu}$ is the effective energy-momentum tensor defined as
\begin{equation}\label{senergy}
S_{\mu\nu}=T_{\mu\nu}-\frac{\kappa\lambda
T}{4\kappa\lambda-1}g_{\mu\nu}.
\end{equation}
Therefore, solutions for the Einstein field equations are also
valid in the Rastall theory, if only we consider $S_{\mu\nu}$ as
the new energy-momentum tensor, for which we have
\begin{eqnarray}\label{scomp}
&S^{0}_{0}&\equiv-\rho^e=-\frac{(3\kappa\lambda-1)\rho+\kappa\lambda(p_r+2p_t)}{4\kappa\lambda-1},\nonumber \\
\nonumber &S^{1}_{1}&\equiv
p^e_r=\frac{(3\kappa\lambda-1)p_r+\kappa\lambda(\rho-2p_t)}{4\kappa\lambda-1},\\
&S^{2}_{2}&=S^{3}_{3}\equiv
p^e_t=\frac{(2\kappa\lambda-1)p_t+\kappa\lambda(\rho-p_r)}{4\kappa\lambda-1}.
\end{eqnarray}
Here, $\rho$, $p_r$ and $p_t$ are the energy density and pressures
corresponding to $T^{\mu}_{\nu}$, respectively. Besides, $\rho^e$,
$p^e_r$ and $p^e_t$ are also the effective energy density and
pressures corresponding to $S^{\mu}_{\nu}$, respectively.
We should note here that $\rho^e$, $p^e_r$ and
$p^e_t$ (effective components) differ from the energy density and
pressures components of original energy-momentum source
($T_{\mu\nu}$). In fact, this effective components have some
geometrical aspects and they help us in comparing energy-momentum
sources satisfying Rastall field equations~(\ref{r1}) with those
satisfying the Einstein field equations~(\ref{ein}). Moreover, as a
desired result, the Einstein field equations are reobtained in the
appropriate limit $\lambda\rightarrow0$. Finally, it is worthwhile
mentioning that the Einstein solutions of $R=0$ is also valid in
this theory \cite{braneb,r0}. One can also use Eq.~(\ref{scomp}) to
see that
\begin{eqnarray}\label{NW}
\rho^e+p_r^e=\rho+p_r,\ \ \rho^e+p_t^e=\rho+p_t,
\end{eqnarray}
meaning that whenever the null and weak energy conditions are
satisfied by the energy-momentum tensor, the effective
energy-momentum tensor will also meet these conditions. It is shown
that, in Rastall's framework, if the weak energy condition is met by
the energy-momentum tensor, then the second law of thermodynamics is
also satisfied by the apparent horizon of the
Friedmann-Lemaitre-Robertson-Walker universe \cite{plb}.
More studies on the thermodynamic properties of this
theory can be found in \cite{ahep,cplb}. In addition, the dominant
energy condition expresses that matter flux should be directed along
the timelike and null geodesics, i.e. $\rho\geq0$ and
$\rho\geq|p_i|$ \cite{poisson}. On the other hand, Raychaudhuri's
equation as well as the Focusing theorem lead to the strong energy
condition ($\rho^e+p_r^e+2p_t^e\geq0$) for the Einstein tensor and
thus $S_{\mu\nu}$ \cite{poisson}. Combining the strong energy
condition with Eq.~(\ref{scomp}), one finds
\begin{eqnarray}\label{SEC}
\rho^e+p_r^e+2p_t^e=\rho+p_r+2p_t+\frac{2\kappa\lambda}{4\kappa\lambda-1}(\rho-p_r).
\end{eqnarray}
Therefore, $\rho+p_r+2p_t\geq0$ if
$\rho^e+p_r^e+2p_t^e\geq\frac{2\kappa\lambda}{4\kappa\lambda-1}(\rho-p_r)$.
Besides, since the time-time component of the Ricci tensor
($R_{00}$) should meet the Newtonian limit
\cite{rastall,cosmos3,smr}, we get
\begin{eqnarray}\label{kappa}
\frac{\kappa}{4\kappa\lambda-1}(3\kappa\lambda-\frac{1}{2})=\kappa_G,
\end{eqnarray}
where $\kappa_G=4\pi G$, and therefore, the Einstein coupling
constant ($\kappa=\kappa_E\equiv8\pi G$) is recovered only while
$\lambda=0$ \cite{rastall,cosmos3}. Solving this equation for
$\lambda$, one reaches
\begin{eqnarray}\label{lambda}
\lambda=\frac{\kappa-2\kappa_G}{6\kappa^2-8\kappa\kappa_G}.
\end{eqnarray}
Here, it is useful to note that Eqs.~(\ref{senergy})
and~(\ref{lambda}) indicate that the dimension of $\lambda$ should
be the inverse of that of $\kappa$ which means
$\lambda\kappa=\gamma$, where $\gamma$ is a dimensionless
constant, we call the Rastall dimensionless parameter. Inserting
this result into~(\ref{kappa}), we obtain
\begin{eqnarray}\label{gamma}
\kappa=\frac{8\gamma-2}{6\gamma-1}\kappa_G.
\end{eqnarray}
We finally define the state parameter $w$ and the effective state
parameter $w_{e}$ as
\begin{eqnarray}\label{omega}
w=\frac{p_r}{\rho},
\end{eqnarray}
and
\begin{eqnarray}\label{omegae}
w_{e}=\frac{p^e_r}{\rho^e},
\end{eqnarray}
respectively. One can use Eqs.~(\ref{scomp}) and~(\ref{omega}) to
get
\begin{eqnarray}\label{denf}
\rho&=&\gamma(p_r^e-\rho^e)+2\gamma p_t^e+\rho^e,\\
\nonumber p_r&=&\gamma(\rho^e-p_r^e)-2\gamma p_t^e+p_r^e,\\
\nonumber p_t&=&\gamma(\rho^e-p_r^e)-2\gamma p_t^e+p_t^e,
\end{eqnarray}
and
\begin{eqnarray}\label{omega1}
w=\frac{\gamma(\rho^e-p_r^e)-2\gamma
p_t^e+p_r^e}{\gamma(p_r^e-\rho^e)+2\gamma p_t^e+\rho^e},
\end{eqnarray}
for the components of $T^\mu_\nu$ and the state parameter,
respectively. From now on, for the sake of simplicity, we set
$8\pi G=1$ or equivalently $\kappa_G=\frac{1}{2}$ which leads to
\begin{eqnarray}\label{gamma1}
\kappa=\frac{4\gamma-1}{6\gamma-1},
\end{eqnarray}
where we used Eq.~(\ref{gamma}) to get the this result.
\section{Traversable asymptotically flat wormholes, general remarks}
Consider the general form of traversable static spherically
symmetric wormholes written as
\begin{equation}\label{met}
ds^{2}=-U(r)dt^{2}+\frac{dr^{2}}{1-\frac{b\left(r\right)}{r}}+r^{2}d\Omega^{2},
\end{equation}
in which $b(r)$ and $U(r)$ are called the shape and redshift
functions, respectively. Additionally,
$d\Omega^{2}=d\theta^{2}+\sin^{2}\theta d\phi^{2}$ is the line
element on the two-dimensional sphere with radius $r$. The zeroth
and radial components of Eq.~(\ref{ein}) lead to
\begin{equation}\label{den}
b^{\prime}(r)=\kappa \rho^er^{2},
\end{equation}
and
\begin{equation}\label{pr0}
\frac{U^{\prime} (r)}{U(r)}=\frac{\kappa p^e_r r^{3}
+b(r)}{r\left(r-b(r)\right)},
\end{equation}
respectively. The last equation can be rewritten as
\begin{equation}\label{pr}
\frac{U^{\prime}(r)}{U(r)}=\frac{rw_{e}
b^{\prime}(r)+b(r)}{r\left(r-b(r)\right)},
\end{equation}
by using Eq.~(\ref{omegae}). The final equation that comes from
the $G^2_2$ component, is
\begin{equation}\label{pe1}
p^e_t=p^e_r+\frac{r}{2}\left[(p^e_r)^{\prime}+\left(\rho^e
+p^e_r\right)\frac{U^{\prime}(r)}{2U(r)}\right],
\end{equation}
for which we have also used Eq.~(\ref{pr0}). In the preceding
formulae, the prime sign denotes the derivative with respect to $r$.
Since we are looking for wormhole solutions, the shape function
should satisfy the $b(r_0)=r_0$ condition, in which $r_0$ is the
wormhole throat radius. Besides, in order to avoid singularities
$U(r)$ should be finite and non-zero everywhere \cite{Thorne2}.
Moreover, the asymptotically flat condition implies the
$(1-\frac{b(r)}{r})\rightarrow1$ and $U(r)\rightarrow1$ conditions
for $r\rightarrow\infty$. The later condition leads to the
$1+z=\frac{1}{\sqrt{U(r_1)}}$ relation for the redshift of a photon
which has been emitted at $r_1$ and is observed at infinity. One can
check that, for $\alpha<1$,
$b(r)=r_0+\beta[(\frac{r}{r_0})^{\alpha}-1]$ is a solution which
satisfies both the $b(r_0)=r_0$ and
$(1-\frac{b(r)}{r})\rightarrow1$ conditions \cite{lpr,hrm}.
Therefore, inasmuch as obtain the
$b(r)=r_0+\beta[(\frac{r}{r_0})^{\alpha}-1]$ relation is independent
of the $U(r)$ function, the mentioned shape function is general and
can be employed for every redshift function. Bearing
Eq.~(\ref{den}) in mind, we obtain
$\rho^e(r)=\frac{\alpha\beta}{\kappa
r^3_0}(\frac{r}{r_0})^{\alpha-3}$. Considering the $\phi(r)=r-c$
hypersurface with normal $n_{\alpha}=\partial_{\alpha}\phi(r)$,
simple calculations lead to $n_{\alpha}n^{\alpha}=n_r
n^r=1-\frac{b(c)}{c}$ at $r=c$ meaning that the $r=c$ hypersurface
is null whenever $c=r_0$. Therefore, inasmuch as the
wormhole throat is a null hypersurface, one may expect that a
radiation source ($w=\frac{1}{3}$) should at least satisfy the
throat of traversable wormholes. But, due to the flaring-out
condition, this expectation cannot be satisfied in the framework of
general relativity \cite{cut-paste1}. Although the wormhole throat
is a null hypersurface, it is not a horizon. The latter is due to
this fact that, in order to obtaining wormhole, the redshift
function should be finite and non-zero everywhere \cite{Thorne2},
meaning that the redshift should not be divergent at $r=r_0$.
Therefore, for avoiding horizon at $r=r_0$, we should have
$U(r_0)\neq0$.

Now, using Eqs.~(\ref{den}) and~(\ref{pr0}), one can evaluate the
effective radial pressure and density at throat as
\begin{eqnarray}\label{atth}
p^e_r(r_0)=-\frac{1}{\kappa r_0^2}\ \textmd{and}\
\rho^e(r_0)=\frac{\alpha\beta}{\kappa r_0^3},
\end{eqnarray}
respectively. Finally, since $p^e_r(r)=w_e(r)\rho^e(r)$, we get the
$w_e(r_0)\beta\alpha=-r_0$ condition. The flaring-out condition also
tells us that the shape function should satisfy the
$b^{\prime}(r_0)<1$ condition, where again the prime denotes the
derivative with respect to $r$ \cite{Thorne2}.
Therefore, the flaring-out condition leads to
$\alpha\beta<r_0$. Moreover, since at the throat we have
\cite{Thorne2}
\begin{eqnarray}\label{flar1}
p^e_r(r_0)+\rho^e(r_0)=\rho^e(r_0)(1+w_e(r_0))=\frac{b^{\prime}(r_0)-1}{\kappa
r_0^2},
\end{eqnarray}
the flaring-out condition leads to $p^e_r(r_0)+\rho^e(r_0)<0$ and
$p^e_r(r_0)+\rho^e(r_0)>0$ for $\kappa>0$ and $\kappa<0$,
respectively. In summary, independent of the value of $\kappa$, the
wormhole parameters, including $\alpha$, $\beta$, $r_0$, and
$w_e(r)$ should meet the $w_e(r_0)\beta\alpha=-r_0$ and
$\alpha\beta<r_0$ conditions. Thus, for $\alpha\beta>0$, one can
find that $w_e(r_0)<-1$. Moreover, since we are looking for
asymptotically flat solutions, we have $\alpha<1$. Finally, we
should remind here that the effective components do not represent a
real fluid. Such geometry has been previously studied in the
Einstein and braneworld frameworks \cite{lpr,pr,hrm,hrmc,chack}. In
what follows, we investigate some properties of such geometry as
well as its corresponding energy-momentum source in the Rastall
framework.

Combining Eqs.~(\ref{NW}) and~(\ref{flar1}) with each other, one can easily find
\begin{eqnarray}\label{flar10}
p_r(r_0)+\rho(r_0)=\rho(r_0)\big(1+w(r_0)\big)=\frac{b^{\prime}(r_0)-1}{\kappa r_0^2},
\end{eqnarray}
meaning that if $\gamma$ meets either the $\gamma<\frac{1}{6}$ or
$\frac{1}{4}<\gamma$ condition (or equally $\kappa>0$), then the
flaring-out condition is obtained whenever we have
$\rho(r_0)\big(1+w(r_0)\big)<0$. In this situation, an
energy-momentum source with negative energy density and $-1<\omega$
may support wormhole. In addition, a source with $\omega<-1$ and
positive energy (a phantom source) can also support this geometry.
Moreover, for negative values of $\kappa$ (or equally
$\frac{1}{6}<\gamma<\frac{1}{4}$), one finds that the flaring-out
condition leads to the $\rho(r_0)\big(1+w(r_0)\big)>0$ condition
meaning that a source with positive energy density and state
parameter which satisfies the $-1<\omega$ condition may also support
wormholes. For this case, it is also easy to obtain that a source
with $\omega<-1$ may support this geometry if it meets the
$\rho(r_0)<0$ condition. We should note that although the negative
energy may support wormholes \cite{nm1,nm2}, due to their various
problems, physicists mostly focus on the phantom solutions
\cite{cut-paste1,nm1,nm2}.

\section{Wormholes with constant redshift function}
Now, we consider the $U(r)=1$ case which respects the asymptotically
flat condition and also leads to $z=0$. As we have
previously mentioned, since the $b(r_0)=r_0$ and
$(1-\frac{b(r)}{r})\rightarrow1$ conditions are enough to obtain the
$b(r)=r_0+\beta[(\frac{r}{r_0})^{\alpha}-1]$ relation, we can use
this shape function in order to continue our study. From
Eqs.~(\ref{pr0}) and~(\ref{pe1}), one obtains
\begin{eqnarray}
p^e_r=-\frac{b(r)}{\kappa
r^3}=-\frac{r_0+\beta[(\frac{r}{r_0})^{\alpha}-1]}{\kappa r^3},
\end{eqnarray}
and
\begin{eqnarray}
p^e_t=-\frac{p^e_r+\rho^e}{2},
\end{eqnarray}
respectively, where $\rho^e(r)=\frac{\alpha\beta}{\kappa
r^3_0}(\frac{r}{r_0})^{\alpha-3}$. Therefore, for the effective
state parameter, we reach
\begin{eqnarray}\label{ee}
w_e(r)=-\frac{b(r)}{rb^{\prime}(r)}=-\frac{r_0+\beta[(\frac{r}{r_0})^{\alpha}-1]}{\alpha\beta(\frac{r}{r_0})^{\alpha}},
\end{eqnarray}
which, as a check, leads to $w_e(r_0)=-\frac{r_0}{\alpha\beta}$ at
the wormhole throat. By combining
Eqs.~(\ref{denf}),~(\ref{omega1}) and~(\ref{gamma1}) with the
above results, we find
\begin{eqnarray}\label{source}
\rho&=&\frac{\alpha\beta(1-2\gamma)(6\gamma-1)}{(4\gamma-1) r_0^3}(\frac{r}{r_0})^{\alpha-3}, \\
p_r&=&\frac{6\gamma-1}{4\gamma-1}[\frac{2\alpha\beta\gamma}{
r_0^3}(\frac{r}{r_0})^{\alpha-3}+\frac{\beta-r_0-\beta
(\frac{r}{r_0})^{\alpha}}{r^3}], \nonumber \\
p_t&=&\frac{6\gamma-1}{4\gamma-1}[\frac{\alpha\beta(4\gamma-1)}{2
r_0^3(\frac{r}{r_0})^{3-\alpha}}-\frac{\beta-r_0-\beta
(\frac{r}{r_0})^{\alpha}}{2r^3}],\nonumber
\end{eqnarray}
and
\begin{eqnarray}\label{omega2}
w(r)=\frac{1}{1-2\gamma}[2\gamma-\frac{1}{\alpha}+\frac{\beta-r_0}{\alpha\beta
(\frac{r}{r_0})^\alpha}],
\end{eqnarray}
for the components of $T^\mu_\nu$ and the state parameter,
respectively.
\subsection{The $0\leq\alpha<1$ case}
Eq.~(\ref{omega2}) leads to
\begin{eqnarray}\label{omega21}
w(r_0)=\frac{1}{1-2\gamma}[2\gamma-\frac{r_0}{\alpha\beta}],
\end{eqnarray}
at the wormhole throat. Moreover, since the wormhole throat is a
null hypersurface, one may equal the above state parameter with
that of radiation ($\frac{1}{3}$) and get
\begin{eqnarray}\label{omega212}
\beta=\frac{3r_0}{\alpha(8\gamma-1)}.
\end{eqnarray}
Now, since $\alpha\beta<r_0$, simple calculations
yield $\frac{1}{2}<\gamma$ meaning that $\kappa$ is positive, and
therefore, based on Eq.~(\ref{flar10}), the flaring-out condition is
met if we have $\rho(r_0)<0$.

\begin{figure}[ht]
\centering
\includegraphics[scale=0.4]{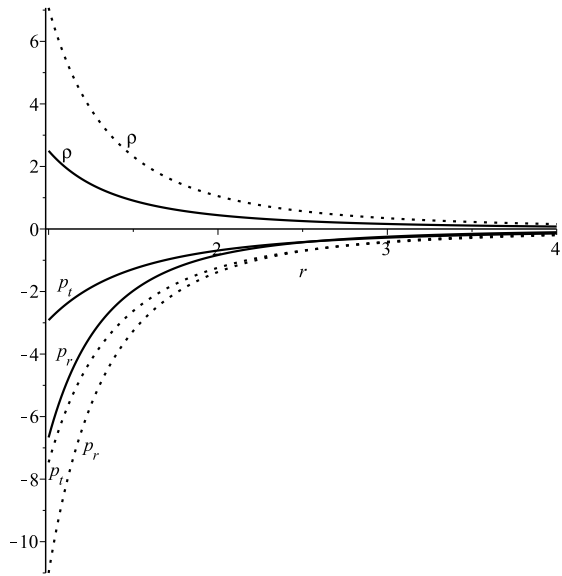}
\caption{\label{en2f} The plot depicts $\rho$, $p_r$ and $p_t$ as
the functions of radius. Solid lines: $\alpha=\frac{1}{2}$,
$\gamma=1$ and $\beta=-3$. Dot lines: $\alpha=\frac{1}{4}$,
$\gamma=2$ and $\beta=-6$.}
\end{figure}

\begin{figure}[ht]
\centering
\includegraphics[scale=0.4]{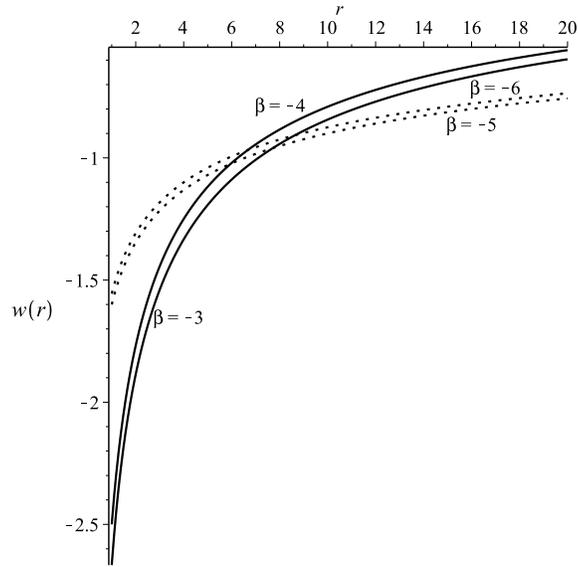}
\caption{\label{werm1ff} The plot depicts $w(r)$ function for some
values of $\gamma$. Solid lines: $\alpha=\frac{1}{2}$ and
$\gamma=1$. Dot lines: $\alpha=\frac{1}{4}$ and $\gamma=2$.}
\end{figure}

As the second example, we look for solutions that
satisfy the $w(r\rightarrow\infty)\rightarrow0$ condition. Applying
the $r\rightarrow\infty$ limit on Eq.~(\ref{omega2}), one can obtain
\begin{eqnarray}\label{omega213}
w=\frac{1}{1-2\gamma}[2\gamma-\frac{1}{\alpha}].
\end{eqnarray}
Bearing the $w(r\rightarrow\infty)\rightarrow0$ condition in mind, a
simple calculation leads to
\begin{eqnarray}\label{omega214}
\alpha=\frac{1}{2\gamma},
\end{eqnarray}
as the mutual relation between $\alpha$ and $\gamma$. Inserting
this result into Eq.~(\ref{omega2}), one obtain
\begin{eqnarray}\label{omega215}
w(r)=\frac{2\gamma(\beta-r_0)}{(1-2\gamma)\beta(\frac{r}{r_0})^{\alpha}}.
\end{eqnarray}
It is useful to note here that $\gamma$ should meet the
$\gamma>\frac{1}{2}$ condition to cover the $0\leq\alpha<1$ case.
Besides, since $\alpha\beta<r_0$,
Eq.~(\ref{omega214}) implies $\beta<2\gamma r_0$. In
Figs.~(\ref{en2f}) and~(\ref{werm1ff}), energy density, the pressure
components and the state parameter are plotted, respectively, in the
exterior of a wormhole with radius $r_0=1$. It is interesting to
note that, unlike the pressure components, energy density is
positive for these solutions. In fact, the positivity of energy
density is due to the $\beta(1-2\gamma)$ term in
Eq.~(\ref{source}) which is positive for $\gamma>\frac{1}{2}$, while
$\beta<0<\frac{r_0}{\alpha}$. For these solutions, as it is clear
from Eq.~(\ref{omega215}) and Fig.~(\ref{werm1ff}), we have
$w(r)\rightarrow0$ at the $r\rightarrow\infty$ limit. The weak
energy condition is also violated by the plotted cases.

Now, let us consider situation in which
$w(r\rightarrow\infty)\rightarrow\eta$, where $\eta$ is an arbitrary
constant. Therefore, from Eq.~(\ref{omega213}), we reach at
\begin{eqnarray}\label{x1}
\alpha=\frac{2\gamma}{1-2\gamma\eta(1-2\gamma)},
\end{eqnarray}
combined with Eq.~(\ref{source}) to find
\begin{eqnarray}\label{x2}
\rho&=&\frac{2\gamma\beta(1-2\gamma)(6\gamma-1)}{[1-2\gamma\eta(1-2\gamma)](4\gamma-1) r_0^3}(\frac{r}{r_0})^{\frac{2\gamma}{1-2\gamma\eta(1-2\gamma)}-3}, \\
p_r&=&\frac{6\gamma-1}{4\gamma-1}[\frac{4\gamma^2\beta}{[1-2\gamma\eta(1-2\gamma)]
r_0^3}(\frac{r}{r_0})^{\frac{2\gamma}{1-2\gamma\eta(1-2\gamma)}-3}+\frac{\beta-r_0-\beta
(\frac{r}{r_0})^{\frac{2\gamma}{1-2\gamma\eta(1-2\gamma)}}}{r^3}], \nonumber \\
p_t&=&\frac{6\gamma-1}{4\gamma-1}[\frac{2\gamma\beta(4\gamma-1)}{2[1-2\gamma\eta(1-2\gamma)]
r_0^3(\frac{r}{r_0})^{3-\frac{2\gamma}{1-2\gamma\eta(1-2\gamma)}}}-\frac{\beta-r_0-\beta
(\frac{r}{r_0})^{\frac{2\gamma}{1-2\gamma\eta(1-2\gamma)}}}{2r^3}].\nonumber
\end{eqnarray}
\begin{figure}[ht]
\centering
\includegraphics[scale=0.4]{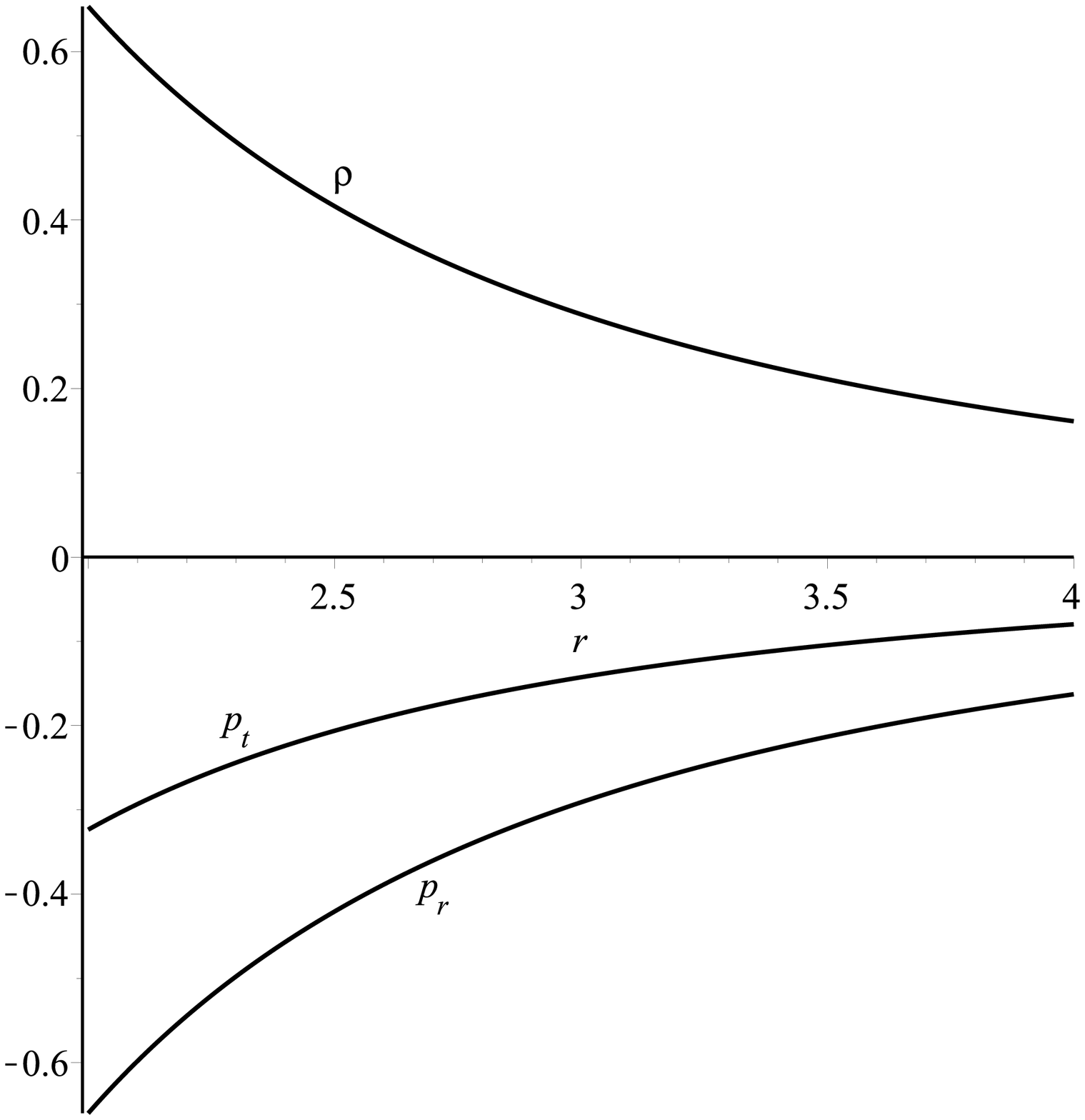}
\caption{\label{xf1} The plot depicts $\rho$, $p_r$ and $p_t$ as the
functions of radius for $\beta=r_0=2$, $\gamma=-0\cdot5$ and
$\eta=-1\cdot01$.}
\end{figure}
As another example, we consider the $\beta=r_0$ case, leading to
$w(r)=\eta$, and find that for $\gamma<\frac{\alpha-1}{4\alpha}<0$
(and thus $\kappa>0$), energy density is positive and pressure
components are negative. Additionally, one can check to see that the
flaring-out condition is also satisfied in this situation. The
behavior of non-zero components of energy-momentum tensor have been
plotted in Fig.~(\ref{xf1}) for a wormhole with $a=0\cdot98$.
Finally, we should indicate that since $\gamma$ is negative and
$0\leq\alpha<1$, we always have $\eta<-1$ meaning that it is a
phantom solution.
\subsection{The $\alpha<0$ case}\label{w0}
Here, we investigate wormholes with $\alpha\leq0$. From
Eq.~(\ref{omega2}), it is apparent that, in order to have a
non-divergent state parameter at the $r\rightarrow\infty$ limit,
we should have $\beta=r_0$. Therefore, we confine ourselves to the
$\beta=r_0$ case and get
\begin{eqnarray}\label{omega21f}
w=\frac{1}{1-2\gamma}[2\gamma-\frac{1}{\alpha}],
\end{eqnarray}
for the state parameter as a function of $\gamma$ and $\alpha$.
In addition, the $\alpha\beta<r_0$ constraint leads
to $\alpha<1$ and therefore, the flaring-out condition
($\alpha\beta<r_0$) is automatically respected by wormholes of
$\alpha<0$ in Rastall's framework. In this situation,
$\rho^e(r_0)=\frac{\alpha}{\kappa r^2_0}$, and we can also use
Eq.~(\ref{ee}) to reach at $w_e=-\frac{1}{\alpha}$. Therefore, for
$\alpha<0$, the effective state parameter meets the $0<w_e$
condition. Now, using the results obtained from Eq.~(\ref{flar1}),
we find that the flaring-out condition is satisfied. It is due to
this fact that, since $1+w_e>0$, for positive (negative) $\kappa$,
we should have $\rho^e(r_0)<0$ ($\rho^e(r_0)>0$), a result respected
by the $\rho^e(r_0)=\frac{\alpha}{\kappa r^2_0}$ expression.

As the first example, consider the $w=\frac{1}{3}$ case leading to
\begin{eqnarray}\label{omega21f2}
\alpha=\frac{3}{8\gamma-1}.
\end{eqnarray}
It can be combined with the $\alpha<0$ condition to
get $\gamma<\frac{1}{8}$ meaning that $\kappa$ is
positive~(Eq.~\ref{gamma1}), and thus, we have $\rho^e(r_0)<0$ and
$\rho^e(r_0)+p^e(r_0)<0$~(see (Eq.~\ref{flar1})). Therefore,
wormholes with $\alpha<0$ and $\beta=r_0$ may be supported by a
fluid with the same state parameter as that of the radiation source
($w=\frac{1}{3}$) in Rastall theory with $\gamma<\frac{1}{8}$.
Inserting Eq.~(\ref{omega21f2}) into~(\ref{source}) and
using~(\ref{gamma1}), we obtain
\begin{eqnarray}\label{source1}
p_r(r)&=&w\rho(r)=\frac{(6\gamma-1)(1-2\gamma)}{(4\gamma-1)(8\gamma-1)r_0^2}(\frac{r}{r_0})^{\frac{3}{8\gamma-1}-3},\nonumber \\
p_t(r)&=&2\frac{(5\gamma-1)}{(1-2\gamma)}p_r(r).
\end{eqnarray}
Since $\gamma<\frac{1}{8}$, unlike the transverse pressure, energy
density and radial pressure are negative, and therefore, ordinary
energy-momentum sources, which have positive energy density, cannot
support these solutions.

As the second example, we consider the $w=0$ case for which
Eq.~(\ref{omega21f}) leads to $\alpha=\frac{1}{2\gamma}$,
and thus $\gamma<0$ to respect the $\alpha<0$
condition. Additionally, from~(\ref{source}) we get
\begin{eqnarray}\label{source12}
\rho(r)&=&\frac{(6\gamma-1)(1-2\gamma)}{(4\gamma-1)2\gamma r_0^2}(\frac{r}{r_0})^{\frac{1}{2\gamma}-3},\nonumber \\
p_t(r)&=&\frac{(6\gamma-1)}{2(1-2\gamma)}\rho(r).
\end{eqnarray}
It is apparent that $p_t(r)>0$ and energy density is negative for
$\gamma<0$ meaning that a dust source (a source with $\rho>0$ and
$w=0$) cannot support this geometry in Rastall theory.

\begin{figure}[ht]
\centering
\includegraphics[scale=0.4]{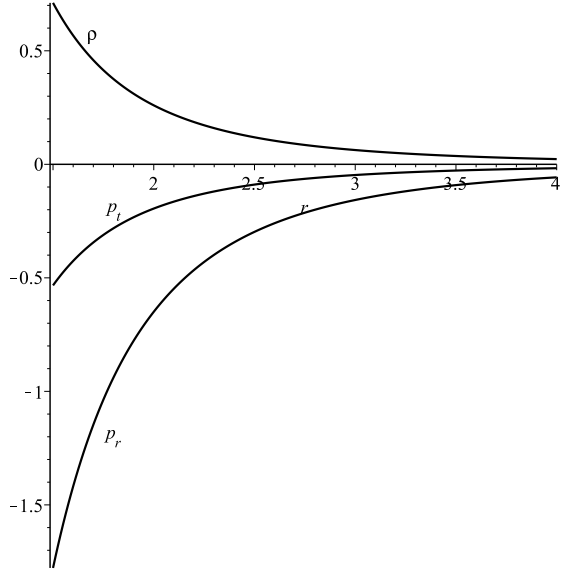}
\caption{\label{cmf2} The plot depicts $\rho$, $p_r$ and $p_t$ as
the functions of radius for $\beta=r_0=\frac{3}{2}$ and
$\alpha=-0\cdot5$.}
\end{figure}

Finally, since a fluid with $w\leq-\frac{2}{3}$ is needed to
describe the current phase of the universe expansion \cite{roos},
we consider the $w=-\frac{5}{6}$ case. Inserting it
into~(\ref{omega21f}), one reaches
\begin{eqnarray}\label{dark}
\alpha=\frac{6}{2\gamma+5},
\end{eqnarray}
and therefore, whenever $\gamma<-\frac{5}{2}$,
leading to $\kappa>0$, the $\alpha<0$ condition will be satisfied.
Combining~(\ref{dark}) and~(\ref{source}), we obtain
\begin{eqnarray}
\rho(r)&=&-\frac{6}{5}p_r(r)=\frac{6(6\gamma-1)(1-2\gamma)}{(4\gamma-1)(2\gamma+5) r_0^2}(\frac{r}{r_0})^{\frac{6}{2\gamma+5}-3},\nonumber \\
p_t(r)&=&\frac{(26\gamma-1)}{12(1-2\gamma)}\rho(r).
\end{eqnarray}
As it is obvious, for $\gamma<-\frac{5}{2}$, we have
$\rho(r)<0$ whenever the pressure components are positive. In
addition, from Eqs.~(\ref{omega21f}) and~(\ref{source}), one finds
\begin{eqnarray}\label{cm1}
\rho&=&\frac{\alpha(1-2\gamma)(6\gamma-1)}{(4\gamma-1) r_0^2}(\frac{r}{r_0})^{\alpha-3}, \\
p_r&=&\frac{(6\gamma-1)(2\gamma\alpha-1)}{(4\gamma-1) r_0^2}(\frac{r}{r_0})^{\alpha-3}, \nonumber \\
p_t&=&\frac{6\gamma-1}{4\gamma-1}[\frac{\alpha(4\gamma-1)}{2
r_0^2(\frac{r}{r_0})^{3-\alpha}}+\frac{1}{2r^2_0}(\frac{r}{r_0})^{\alpha}],\nonumber
\end{eqnarray}
where $\beta=r_0$ has also been used to obtain these results. Since
$\alpha$ is negative, energy density is positive whenever $\gamma$
either meets the $\frac{1}{6}<\gamma<\frac{1}{4}$ or
$\frac{1}{2}<\gamma$ condition. In this situation, $p_r$ and $w$ are
positive (negative) for $\frac{1}{6}<\gamma<\frac{1}{4}$
($\frac{1}{2}<\gamma$). As we have previously seen in
Eq.~(\ref{flar10}), for $\kappa<0$ (or equally
$\frac{1}{6}<\gamma<\frac{1}{4}$), the flaring-out condition is
satisfied if $\rho(r_0)+p_r(r_0)>0$. In addition,
$\rho(r_0)+p_r(r_0)=\frac{(6\gamma-1)(\alpha-1)}{(4\gamma-1)r_0^2}$
which is positive for $\kappa<0$ and negative for
$\frac{1}{2}<\gamma$ (or equally $\kappa>0$), and therefore, based
on Eq.~(\ref{flar10}), the flaring-out condition is obtained.
\begin{figure}[ht]
\centering
\includegraphics[scale=0.4]{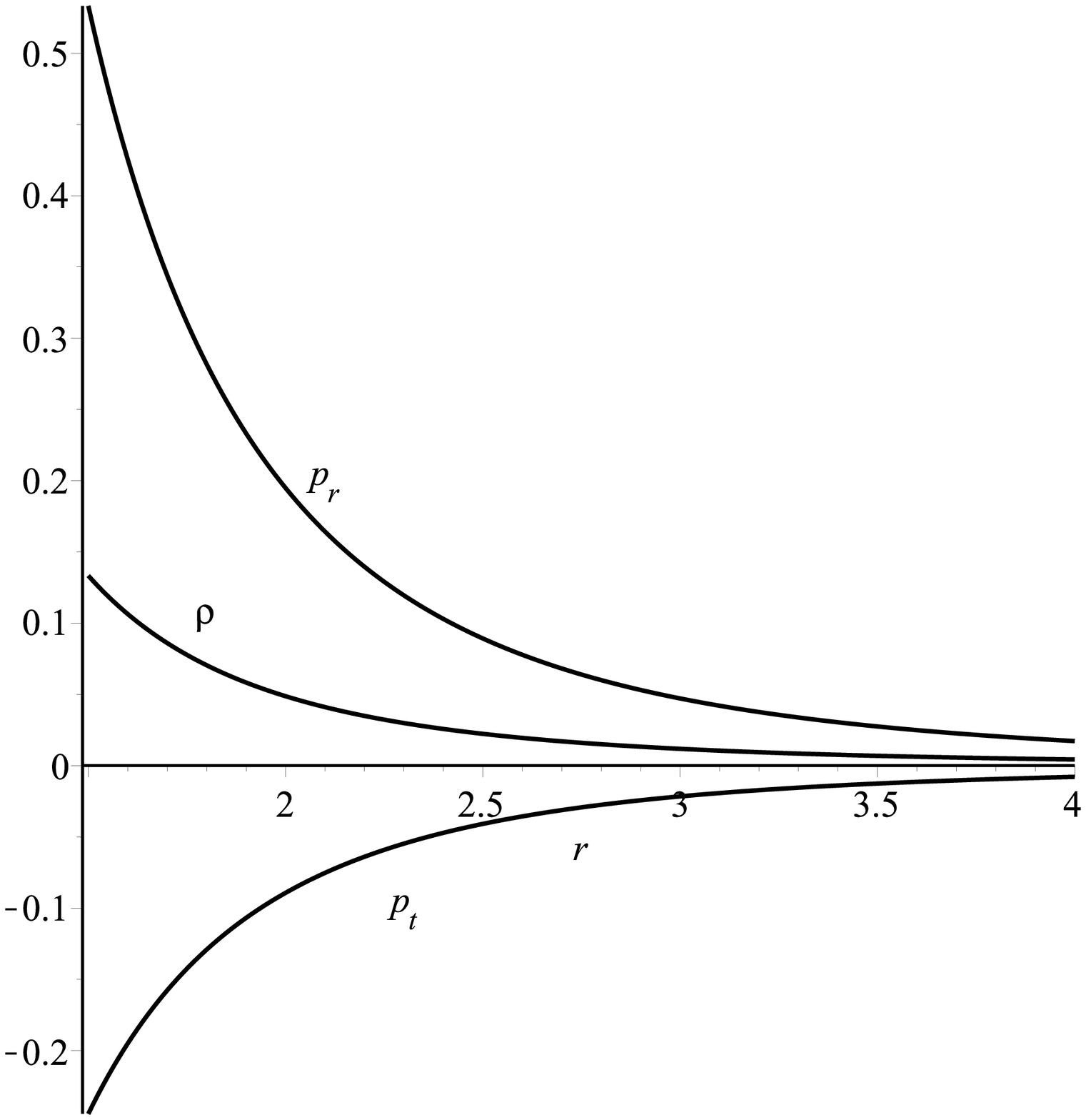}
\caption{\label{cmf1} The plot depicts $\rho$, $p_r$ and $p_t$ as
the functions of radius for $\beta=r_0=\frac{3}{2}$ and
$\alpha=-0\cdot5$.}
\end{figure}
In Figs.~(\ref{cmf1}) and~(\ref{cmf2}), we have plotted the non-zero
components of energy-momentum tensor for $\gamma=\frac{1}{5}$ and
$\gamma=\frac{3}{2}$, respectively. Using Eq.~(\ref{omega2}), one
finds that, since $\alpha$ is negative,
$w=\frac{2\alpha-5}{3\alpha}>\frac{2}{3}$ for $\gamma=\frac{1}{5}$,
and moreover, $w=\frac{1-3\alpha}{2\alpha}<-\frac{3}{2}$ (phantom
solution), whenever $\gamma=\frac{3}{2}$.
Finally, it is worthwhile remembering that since for
$\gamma=\frac{1}{5}$ ($\gamma=\frac{3}{2}$) we have $\kappa<0$
($\kappa>0$), the flaring-out condition leads to the
$\rho(r_0)+p_r(r_0)>0$ ($\rho(r_0)+p_r(r_0)<0$) condition which is
met based on the results obtained from the
$\rho(r_0)+p_r(r_0)=\frac{(6\gamma-1)(\alpha-1)}{(4\gamma-1)r_0^2}$
relation.

\section{Wormholes with constant effective state parameter}
Here, we investigate some properties of a source that supports
asymptotically flat wormholes with constant effective state
parameter and $b(r)=r_0+\beta[(\frac{r}{r_0})^{\alpha}-1]$, in the
Rastall framework. Since the effective state parameter is
constant, $w_e(r_0)\alpha\beta=-r_0$ is reduced to
$w_e\alpha\beta=-r_0$ which leads to
$w_e=-\frac{r_0}{\alpha\beta}$. Using this result and
Eq.~(\ref{pr}), we get
\begin{eqnarray}\label{shap}
U(r)=C\exp(\int\frac{(\frac{r}{r_0})^{\alpha}(\beta-r_0)+r_0-\beta}{r(r-r_0+\beta(1-(\frac{r}{r_0})^{\alpha}))}dr),
\end{eqnarray}
where $C$ is an integration constant and may be found by
asymptotically flat condition. Combining Eqs.~(\ref{omegae})
and~(\ref{omega1}) with each other, one reaches
\begin{eqnarray}\label{omegaf1}
w(r)=\frac{w_e(1-\gamma)+\gamma(1-2w_e^t)}{\gamma(w_e+2w_e^t)+1-\gamma},
\end{eqnarray}
where $w_e^t=\frac{p^e_t}{\rho^e}$, for the state parameter. In
addition, from Eqs.~(\ref{pe1}) and~(\ref{pr}) we obtain
\begin{eqnarray}\label{omegaft1}
w_e^t=w_e+\frac{r}{2}[\frac{w_e(\rho^e)^{\prime}}{\rho^e}+(1+w_e)\frac{b+r
w_e b^{\prime}}{2r(r-b)}].
\end{eqnarray}

For example, inserting $\alpha=-1$ into
Eq.~(\ref{shap}), one reaches at
\begin{eqnarray}\label{shap01}
U(r)=C(r-r_0)^{\frac{r_0-1}{r_0+\beta}}(r+\beta)^{\frac{\beta^2-r_0}{\beta(\beta+r_0)}}r^{\frac{1-\beta}{\beta}}.
\end{eqnarray}
Therefore, in order to avoid the $r=r_0$ singularity, we should have
$r_0=1$ and $\beta>-1=-r_0$ which lead to
\begin{eqnarray}\label{shap02}
U(r)=(\frac{r+\beta}{r})^{\frac{\beta-1}{\beta}},
\end{eqnarray}
where we also considered the asymptotically flat condition to get
the above result.

From now on, for the sake of simplicity, we only focus on the
$r_0=1$ case yielding $w_e\alpha\beta=-1$ and
$\alpha\beta<1$. Inserting Eq.~(\ref{den}) and
$b(r)=1+\beta[r^{\alpha}-1]$ into Eq.~(\ref{omegaf1}) and using the
$w_e\alpha\beta=-1$ condition, we finally get
\begin{eqnarray}\label{omegaf2}
w(r)=\frac{\frac{2w_e(1-\gamma\alpha)}{(w_e+1)A(r)}+\gamma(\frac{2}{(w_e+1)A(r)}+1)}
{\gamma(\frac{2w_e\alpha}{(w_e+1)A(r)}+1)+\frac{2(1-\gamma)}{(w_e+1)A(r)}},
\end{eqnarray}
in which
$A(r)=\frac{(\beta-1)r^{\alpha}+1-\beta}{r-1+\beta(1-r^{\alpha})}$
and in the $r\rightarrow\infty$ limit,
$w\rightarrow\frac{w_e+\gamma(1-w_e\alpha)}{1-\gamma(1-w_e\alpha)}$,
for $\alpha<1$. In addition, in the $\kappa=1$ limit, which leads
to $\gamma=0$ and thus $\lambda=0$~(\ref{gamma1}), the result of
Einstein theory, i.e. $w(r)\rightarrow w_e$ is reobtained
\cite{lpr}.
\subsection{Solutions with asymptotically zero state parameter}
For solutions in which state parameter vanishes asymptotically
($w(r\rightarrow\infty)\rightarrow0$), we get
\begin{eqnarray}\label{cond1}
w_e=\frac{\gamma}{\gamma\alpha-1},
\end{eqnarray}
for the effective state parameter as a function of $\gamma$ and
$\alpha$. Besides, since $w_e\alpha\beta=-1$, one finds
\begin{eqnarray}\label{betaf1}
\beta=\frac{1-\gamma\alpha}{\gamma\alpha},
\end{eqnarray}
for the $\beta$ parameter. Bearing the
$\alpha\beta<1$ condition in mind, we can use the
above equation to obtain
\begin{eqnarray}\label{gammaf1}
\frac{1-\gamma}{\gamma}<\alpha,
\end{eqnarray}
available if $\gamma$ meets either $\gamma<0$ or
$\frac{1}{2}<\gamma$. In this situation, Eq.~(\ref{gammaf1}) gives a
lower bound for $\alpha$, and therefore, we should have
$\frac{1-\gamma}{\gamma}<\alpha<1$ to meet the the asymptotically
flat condition ($\alpha<1$). Inserting~(\ref{betaf1})
and~(\ref{cond1}) into~(\ref{omegaf2}), one can obtain
\begin{eqnarray}\label{omegaf22}
w(r)=\frac{\frac{2\gamma(\gamma\alpha-1)}{B(r)(\gamma(\alpha+1)-1)}+
\gamma(\frac{2(\gamma\alpha-1)}{B(r)(\gamma(\alpha+1)-1)}+1)}{\gamma(\frac{2\gamma\alpha}{(\gamma(1+\alpha)-1)B(r)}+1)
+\frac{2(1-\gamma)(\gamma\alpha-1)}{(\gamma(1+\alpha)-1)B(r)}},
\end{eqnarray}
in which
$B(r)=\frac{(1-2\gamma\alpha)r^{\alpha}+2\gamma\alpha-1}{(r-1)\gamma\alpha+(1-\gamma\alpha)(1-r^{\alpha})}$.

For the $\alpha=-1$ case, using
Eqs.~(\ref{betaf1}),~(\ref{den}) and~(\ref{flar1}), one can easily
see that the flaring-out condition is satisfied when $\gamma<0$
leading to $\kappa>0$.
\begin{figure}[ht]
\centering
\includegraphics[scale=0.4]{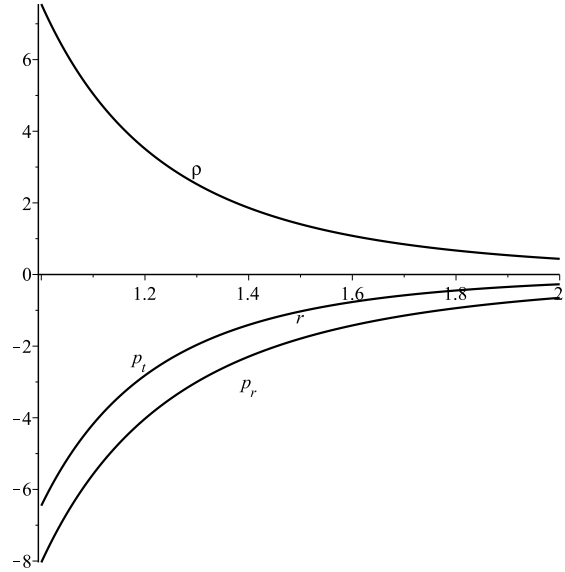}
\caption{\label{asmc1} The plot depicts $\rho$, $p_r$ and $p_t$ as
the functions of radius for $\beta=\frac{1}{w_e}=-\frac{2}{3}$ and
$\gamma=3\alpha=-3$.}
\end{figure}
In fact, for $-1<\gamma<0$, although $-1<w_e<0$, we have $\rho^e<0$
and therefore, based on Eq.~(\ref{flar1}), the flaring-out condition
is met. In addition, for $\gamma<-1$, we have $w_e<-1$ and
$\rho^e>0$ meaning that the flaring-out condition is satisfied.
Considering a Rastall theory of $\gamma=-3$, we have plotted this
case for $\beta=\frac{1}{w_e}=-\frac{2}{3}$, which is a phantom
solution, in Fig.~(\ref{asmc1}).

\subsection{Solutions with asymptotically radiation state parameter}

In order to get solutions with asymptotically radiation state
parameter, following the above recipe, we get
\begin{eqnarray}\label{cond1f}
w_e=\frac{1-4\gamma}{3-4\gamma\alpha},
\end{eqnarray}
for the effective state parameter. In obtaining this
result we used the ($w(r\rightarrow\infty)\rightarrow\frac{1}{3}$)
condition. Moreover, combining Eq.~(\ref{cond1f}) with the
$\omega_e\alpha\beta=-1$ condition, we reach at
\begin{eqnarray}\label{betaff1}
\beta=\frac{4\gamma\alpha-3}{\alpha(1-4\gamma)}.
\end{eqnarray}
Now, one can combine these equations with Eqs.~(\ref{denf})
and~(\ref{omegaf2}) to get the non-zero components of
energy-momentum tensor as well as the state parameter, respectively.
\begin{figure}[ht]
\centering
\includegraphics[scale=0.4]{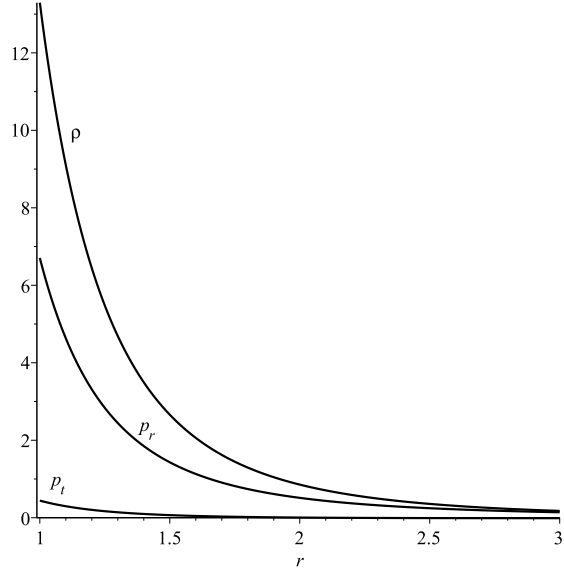}
\caption{\label{asrc1} The plot depicts $\rho$, $p_r$ and $p_t$ as
the functions of radius for $\beta=19$, $\gamma=\frac{1}{5}$, and
$\alpha=-1$.}
\end{figure}
The $\beta=19$, $\gamma=\frac{1}{5}$, $\alpha=-1$ has been plotted
in Fig.~(\ref{asrc1}). Here, since $\gamma=\frac{1}{5}$, we have
$\kappa=-1<0$ and therefore, based on Eq.~(\ref{flar10}), the
flaring-out condition is met if $\rho(r_0)+p_r(r_0)>0$, a condition
obtained by this case. Therefore, for these solutions, it is
theoretically possible to respect the flaring-out condition and
energy conditions simultaneously.
\subsection{The $\beta=r_0=1$ case}
Inserting $\beta=r_0=1$ into~(\ref{omegaf2}), one can obtain
\begin{eqnarray}\label{omegab}
w=\frac{2w_e(1-\gamma\alpha)+2\gamma}{2w_e\alpha\gamma+2(1-\gamma)},
\end{eqnarray}
for the state parameter. Moreover, since $\beta=1$, the $w_e
\beta\alpha=-1$ condition leads to $w_e=-\frac{1}{\alpha}$. By
substituting this result into the last equation, we arrive at
\begin{eqnarray}\label{omegab2}
w=\frac{2\gamma\alpha-1}{(1-2\gamma)\alpha},
\end{eqnarray}
for the state parameter. It is also obvious that, since
$\beta=r_0=1$, the $\alpha\beta<1$ condition is satisfied
whenever $\alpha<1$. Additionally, from Eq.~(\ref{shap}), we
obtain
\begin{eqnarray}\label{shap2}
U(r)=C.
\end{eqnarray}
Therefore, the asymptotically flat condition implies $C=1$ and
finally, one gets
\begin{eqnarray}\label{prop1}
\rho(r)&=&\frac{\alpha r^{\alpha-3}(6\gamma-1)}{4\gamma-1}(1-2\gamma),\nonumber \\
p_r(r)&=&\frac{r^{\alpha-3}(6\gamma-1)}{4\gamma-1}(2\gamma\alpha-1), \\
p_t(r)&=&\frac{r^{\alpha-3}(6\gamma-1)}{2(4\gamma-1)}(\alpha(4\gamma-1)+1).\nonumber
\end{eqnarray}
Here, we should note that although these results are similar to
those obtained in Sec.~(\ref{w0}), there is a key difference
between these results and those addressed in~(\ref{w0}). While in
Sec.~(\ref{w0}), we have $\beta=r_0$ where $r_0$ is an arbitrary
quantity, here, $\beta=r_0$ and $r_0$ must be equal to $1$.

As the first example, consider the $w=0$ case
leading to $\alpha=\frac{1}{2\gamma}$, $p_r=0$, and finally
$p_t(r)=\frac{6\gamma-1}{2(1-2\gamma)}\rho$, where
$\rho(r)=\frac{(6\gamma-1)(1-2\gamma)}{2\gamma(4\gamma-1)}r^{\frac{1-6\gamma}{2\gamma}}$.
For these solutions, since the asymptotically flat condition implies
$\alpha<1$, the Rastall dimensionless parameter should meet the
$\frac{1}{2}<\gamma$ condition meaning that the energy density is
negative. Therefore, we do not focus on this case further.

As the second example, we consider the $w=\frac{1}{3}$ case. Simple
calculations yield $\alpha=\frac{3}{8\gamma-1}$,
$\rho(r)=\frac{p_r(r)}{3}=\frac{3(6\gamma-1)(1-2\gamma)}{(4\gamma-1)(8\gamma-1)}r^{\frac{6(1-4\gamma)}{8\gamma-1}}$
and
$p_t(r)=\frac{2(6\gamma-1)(5\gamma-1)}{(4\gamma-1)(8\gamma-1)}r^{\frac{6(1-4\gamma)}{8\gamma-1}}$.
For these solutions, energy density and radial
pressure are positive whenever $\frac{1}{4}<\gamma<\frac{1}{2}$.
But, for these values of $\gamma$, we have $\kappa>0$ and therefore,
based on Eq.~(\ref{flar10}), the flaring-out condition is not
satisfied.

Using Eq.~(\ref{omegab2}), one can easily find that
phantom solutions ($w<-1$) may be obtained whenever one of the below
conditions is met:
\begin{enumerate}
\item $\gamma>\frac{1}{2}$ and $\alpha<0$.
\item $\gamma<\frac{1}{2}$ and $0<\alpha<1$.
\end{enumerate}

For the first case, we have $\kappa>0$ and energy
density is positive. Moreover, from Eq.~(\ref{flar10}), it is far
from apparent that the flaring-out condition is also satisfied.
\begin{figure}[ht]
\centering
\includegraphics[scale=0.4]{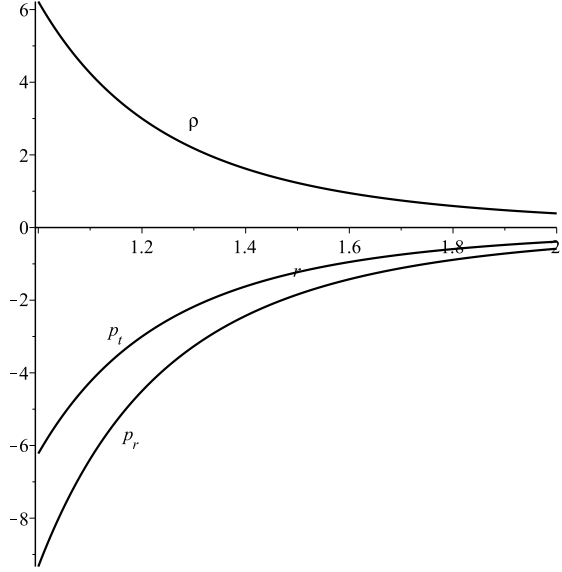}
\caption{\label{cmff1} Here, $\alpha=-1$, $\gamma=\frac{5}{2}$, and
thus $w=-\frac{3}{2}$.}
\end{figure}
In Fig.~(\ref{cmff1}), non-zero components of energy-momentum tensor
have been plotted for $\alpha=-1$ and $\gamma=\frac{5}{2}$. It is
obvious that the weak energy condition is violated by this phantom
solution.

On the other hand, positive energy density is
obtainable in the second case if $\gamma$ either meets
$\gamma<\frac{1}{6}$ or $\frac{1}{4}<\gamma<\frac{1}{2}$ again
leading to  $\kappa>0$ and $p_r(r)<0$.
\begin{figure}[ht]
\centering
\includegraphics[scale=0.4]{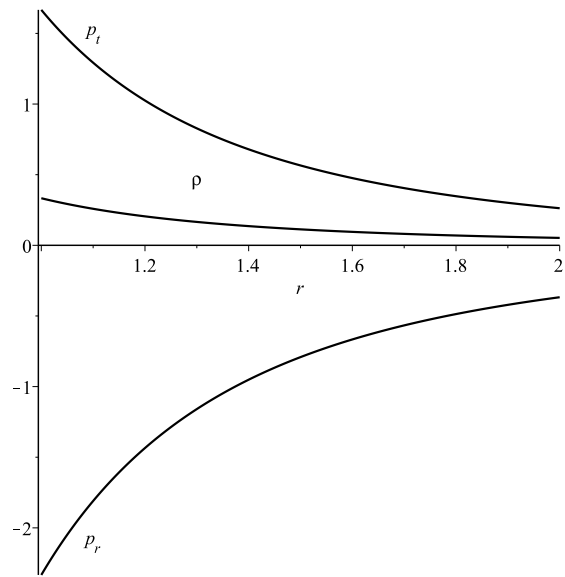}
\caption{\label{cmff2} Here, $\gamma=\alpha=\frac{1}{3}$.}
\end{figure}
The case of $\gamma=\alpha=\frac{1}{3}$ has been depicted in
Fig.~(\ref{cmff2}). As it is apparent, although the transverse
pressure is positive for this solution, since $\rho+p_r<0$, the weak
energy condition is not met by this solution.

\section{Conclusion}
After referring to the Rastall theory, we defined the Rastall
dimensionless parameter ($\gamma$) helping us in simplifying the
calculations. In fact, regarding the Newtonian limit, one can find
a relation between Rastall gravitational coupling constant
($\kappa$) and the Rastall parameter ($\lambda$) on one hand, and
the Newtonian gravitational coupling constant ($\kappa_G$) on the
other hand. Indeed, $\kappa$ and $\lambda$ are unknowns in this
theory and they are only constrained by the Newtonian limit.
Therefore, by finding a suitable value for $\gamma$ and using the
results of Newtonian limit, one can obtain both $\kappa$ and
$\lambda$ parameters. It is also obvious that $\lambda=0$ case
leads to $\gamma=0$ and thus the Einstein field equations are
recovered.

Thereinafter, we considered a general form for the shape function of
traversable asymptotically flat wormholes and studied some cases.
Our results indicate that phantom solutions can be
supported by this theory. Moreover, we found out that, depending on
the value of $\gamma$ and thus $\kappa$, traversable wormholes may
meet both the flaring-out condition and energy conditions in the
Rastall theory. Therefore, our study shows that a non-minimal
coupling between curvature and matter fields may theoretically
support traversable wormholes satisfying energy conditions. In
addition, we found that the wormhole parameters ($\alpha$ and
$\beta$) are affected by the Rastall dimensionless parameter as well
as the assumed primary conditions such as the asymptotically zero-
or radiation-like state parameter. Moreover, we studied wormholes of
$w_e=constant$ and investigated the properties of the
energy-momentum source supporting the geometry in some cases,
including solutions with asymptotically dust- or radiation-like
state parameter, as well as the solutions with constant state
parameter while $\beta=r_0=1$. We also investigated the possibility
of supporting such geometries by a source of $w\leq-\frac{2}{3}$.

Finally, it is important to study the effects of the Rastall
hypothesis on the stability conditions of wormholes. Moreover, the
study of charged wormhole structures in the Rastall framework is
also interesting. It is also worthwhile to investigate the
particle geodesics in the context of obtained wormhole geometry.
We leave these subjects for the future works.

\acknowledgments{We are grateful to the respected referees for their valuable comments. The work of H. Moradpour has been supported
financially by Research Institute for Astronomy \& Astrophysics of
Maragha (RIAAM)} 


\begin{thebibliography}{}
\bibitem{Wheeler1} J. A. Wheeler, Phys. Rev. {\bf97}, 511 (1995).
\bibitem{Wheeler2} J. A. Wheeler, Ann. Phys. {\bf2}, 604 (1957).
\bibitem{Thorne1} M. S. Morris, K. S. Thorne and U. Yurtsewer, Phys. Rev. Lett. {\bf61}, 1446 (1988).
\bibitem{Thorne2} M. S. Morris, K. S. Thorne, Am. J. Phys. {\bf56}, 395 (1988).
\bibitem{Visser1} M. Visser, Phys. Rev. D {\bf39}, 3182 (1989).
\bibitem{Sushkov} S. V. Sushkov, Phys. Rev. D  {\bf71}, 043520 (2005).
\bibitem{Lobo2005} F. S. N. Lobo, Phys. Rev. D {\bf71}, 084011 (2005).
\bibitem{Lobo20051} F. S. N. Lobo, Phys. Rev. D {\bf71}, 124022 (2005).
\bibitem{Hayward} S. A. Hayward, Int. J. Mod. Phys. D {\bf8}, 373 (1999).
\bibitem{Kardashev} N. S. Kardashev, I. D. Novikov and A. A. Shatskiy, Int. J. Mod. Phys. D {\bf16}, 909 (2007).
\bibitem{Kuhfittig} P. K. F. Kuhfittig, Schol. Res. Exch. 296158 (2008).
\bibitem{Sushkov-Zaslavskii} S. V. Sushkov and O. B. Zaslavskii, Phys. Rev. D {\bf79}, 067502 (2009).
\bibitem{Cai1} K. K. Nandi, Y. Z. Zhang and R. G. Cai, [arXiv:gr-qc/0409085].
\bibitem{Cai2} K. K. Nandi, Y. Z. Zhang, R. G. Cai, A. Panchenko, Phys. Rev. D \textbf{79}, 024011 (2009).
\bibitem{ref1} M. A. Aïnou, JCAP. \textbf{07}, 037 (2015).
\bibitem{prds} T. Harko, F. S. N. Lobo, M. K. Mak and S. V. Sushkov, Phys. Rev. D {\bf87}, 067504 (2013).
\bibitem{prds1} R. A. El-Nabulsi, Can. Jour. Phys. DOI: 10.1139/cjp-2017-0109 (2017).
\bibitem{braneb} K. A. Bronnikov and Sung-Won Kim, Phys. Rev. D {\bf67}, 064027 (2003).
\bibitem{Lobo2} F. S. N. Lobo, Phys. Rev. D {\bf75}, 064027 (2007).
\bibitem{Lobo1} F. S. N. Lobo,  Class. Quant. Grav. {\bf25}, 175006 (2008).
\bibitem{Oliveira} F. S. N. Lobo and M. A. Oliveira, Phys. Rev. D {\bf80}, 104012 (2009).
\bibitem{Sajadi} S. N. Sajadi and N. Riazi, Prog. Theor. Phys. {\bf126}, 753 (2011).
\bibitem{Garcia2} N. M. Garcia and F. S. N. Lobo, Phys. Rev. D {\bf82}, 104018 (2010).
\bibitem{Garcia3} N. M. Garcia and F. S. N. Lobo, Class. Quant. Grav. {\bf28}, 085018 (2011).
\bibitem{loboaip} F. S. N. Lobo, AIP. Conf. Proc. {\bf1458}, 447 (2011).



\bibitem{cut-paste1} M. Visser, \emph{Lorentzian Wormholes}, (AIP Press, Woodbury, NY, USA 1996).
\bibitem{cut-paste2} F. S. N. Lobo and P. Crawford, Class. Quant. Grav. \textbf{21}, 391 (2004).
\bibitem{cut-paste3} M. G. Richarte and C. Simeone, Phys. Rev. D \textbf{76}, 087502 (2007); Erratum Phys. Rev. D \textbf{77}, 089903 (2008).
\bibitem{cut-paste4} G. A. S. Dias and J. P. S. Lemos, Phys. Rev. D \textbf{82}, 084023 (2010).
\bibitem{cut-paste44} S. Habib Mazharimousavi, M. Halilsoy and Z. Amirabi, Phys. Lett. A \textbf{375}, 3649 (2011).
\bibitem{cut-paste5} P. Kanti, B. Kleihaus, and J. Kunz, Phys. Rev. D \textbf{85}, 044007 (2012).
\bibitem{cut-paste6} M. H. Dehghani and S. H. Hendi, Gen. Rel. Grav. \textbf{41}, 1853 (2009).
\bibitem{cut-paste7} S. H. Hendi, J. Math. Phys. \textbf{52}, 042502 (2011).
\bibitem{cut-paste8} S. H. Hendi, Prog. Theor. Phys. \textbf{127}, 907 (2012).
\bibitem{cut-paste9} S. H. Hendi, Adv. High Energy Phys. \textbf{2014}, 697863 (2014).
\bibitem{cut-paste10} S. Habib Mazharimousavi, M. Halilsoy, Eur. Phys. J. C \textbf{75}, 271 (2015).
\bibitem{cut-paste11} A. \"{O}vg\"{u}n, [arXiv:1610.08118].
\bibitem{cut-paste12} A. \"{O}vg\"{u}n and I. Sakalli, Theor. Math. Phys. \textbf{190},
120 (2017).


\bibitem{cmc} T. Koivisto, Class. Quant. Grav. {\bf23}, 4289 (2006).
\bibitem{cmc1} O. Bertolami, C. G. Boehmer, T. Harko and F. S. N. Lobo, Phys. Rev. D {\bf75}, 104016 (2007).
\bibitem{cmc2} T. Harko and F. S. N. Lobo, Galaxies, {\bf2}, 410 (2014).
\bibitem{motiv2} L. Parker, Phys. Rev. D {\bf3}, 346 (1971); D {\bf3}, 2546 (1971).
\bibitem{motiv1} G. W. Gibbons and S. W. Hawking, Phys. Rev. D {\bf15}, 2738 (1977).
\bibitem{motiv4}  N. D. Birrell, P. C. W. Davies, \textit{Quantum Fields in Curved Space} (Cambridge University Press, Cambridge, 1982).
\bibitem{motiv3} L. H. Ford, Phys. Rev. D {\bf35}, 2955 (1987).
\bibitem{rw1} Ph. Brax, C. van de Bruck, A. Davis, arXiv:0706.1024 (2007).
\bibitem{rastall} P. Rastall, Phys. Rev. D {\bf 6}, 3357 (1972).
\bibitem{prd} C. E. M. Batista, M. H. Daouda, J. C. Fabris, O. F. Piattella, D. C. Rodrigues, Phys. Rev. D {\bf85}, 084008 (2012).
\bibitem{smal} L. L. Smalley, Il Nuovo Cimento B, {\bf 80}, 42 (1984).
\bibitem{cosmos3} M. Capone, V. F. Cardone, M. L. Ruggiero, Journal of Physics: Conference Series, {\bf222}, 012012 (2010).
\bibitem{al1} A. S. Al-Rawaf, O. M. Taha, Phys. Lett. B 366, 69 (1996).
\bibitem{al3} J. C. Fabris, R. Kerner, J. Tossa, Int. J. Mod. Phys. D 9, 111 (2000).
\bibitem{al2} A. S. Al-Rawaf, O. M. Taha, Gen. Rel. Grav. 28, 935 (1996).
\bibitem{al6} A. S. Al-Rawaf, Int. J. Mod. Phys. D 14, 1941 (2005).
\bibitem{al5} V. Majernik, Gen. Rel. Grav. 35, 1007 (2003).
\bibitem{al55} A. I. Arbab, JCAP. 05, 008 (2003).
\bibitem{al4} A.-M. M. Abdel-Rahman, Astrophys. Space Sci. 278, 383 (2001).
\bibitem{al7} A.-M. M. Abdel-Rahman, M. H. A. Hashim, Astrophys. Space Sci. 298, 519 (2005).
\bibitem{arXiv:1407.3849} E. R. Bezerra de Mello, J. C. Fabris and B.
Hartmann, [arXiv:1407.3849].
\bibitem{Godel} A. F. Santos and S. C. Ulhoa, [arXiv:1407.4322].
\bibitem{r0} K. A. Bronnikov, J. C. Fabris, O. F. Piattella and E. C. Santos, Gen. Relativ. Gravit. {\bf48}, 162 (2016).
\bibitem{plb} H. Moradpour, Phys. Lett. B {\bf757}, 187 (2016)[arXiv:1601.04529].
\bibitem{ahep} H. Moradpour, I. G. Salako, AHEP, 3492796 (2016).
\bibitem{cplb} F. F. Yuan and P. Huang, Class. Quant. Grav. 34, 077001(2017).
\bibitem{poisson} E. Poisson, \textit{A Relativist's Toolkit} (Cambridge University Press, Cambridge, 2004).
\bibitem{smr} A. Sheykhi, H. Moradpour and N. Riazi, Gen. Rel. Grav. {\bf45}, 1033 (2013).
\bibitem{lpr} F. S. N. Lobo, F. Parsaei and N. Riazi, Phys. Rev. D {\bf87}, 084030 (2013).
\bibitem{hrm} Y. Heydarzade, N. Riazi and H. Moradpour, Can. J. Phys {\bf93}, 12 (2015).
\bibitem{pr} F. Parsaei and N. Riazi, Phys. Rev. D {\bf91}, 024015 (2015).
\bibitem{hrmc} D. A. Tretyakova, B. N. Latosh and S. O. Alexeyev. Class. Quantum Grav. {\bf32}, 185002 (2015).
\bibitem{chack} F. Rahaman, M. Kalam, M. Sarker and S. Chakraborty, Acta Phys. Pol. B {\bf40}, 25 (2009).
\bibitem{nm1} S. Hawking, \textit{A Brief History of Time} (Bantam Dell Publishing Group, 1988).
\bibitem{nm2} L. H. Ford and T. A. Roman, Scientific American 282, 46 (2000).
\bibitem{roos} M. Roos, \textit{Introduction to Cosmology} (John Wiley and Sons, UK, 2003).
\end{thebibliography}
\end{document}